\documentclass[11pt]{article}
\usepackage{amsmath}
\usepackage{amssymb}
\usepackage{cite}
\usepackage{txfonts} 
\usepackage{graphicx}
\usepackage{epstopdf}
\usepackage{multicol}
\usepackage{float}
\usepackage{color}

\title{Explosion Dynamics of Parametrized Spherically Symmetric Core-Collapse Supernova Simulations}

\author{Kevin \textsc{Ebinger}$^{1}$, Sanjana \textsc{Sinha}$^{2}$, Carla \textsc{Fr\"ohlich}$^{2}$, Albino \textsc{Perego}$^{3}$, Matthias \textsc{Hempel}$^{1}$,\\ Marius \textsc{Eichler}$^{2}$, Jordi \textsc{Casanova}$^{2}$, Matthias \textsc{Liebend\"orfer}$^{1}$, and Friedrich-Karl \textsc{Thielemann}$^{1}$
\\
\\ 
\small $^{1}$ Department f\"ur Physik, Universit\"at Basel, CH-4056 Basel, Switzerland
\\ \small $^{2}$Department of Physics, North Carolina State University, Raleigh, NC, 29695-8202, USA \\
\small $^{3}$Institut f\"ur Kernphysik, Technische Universit\"at Darmstadt, D-64289 Darmstadt, Germany
}

\date{August 20, 2016}

\begin{document}

\maketitle
\abstract{We report on a method, PUSH, for triggering core-collapse supernova (CCSN) explosions of massive stars in spherical symmetry. This method provides a framework to study many important aspects of core collapse supernovae: the effects of the shock passage through the star, explosive supernova nucleosynthesis and the progenitor-remnant connection. Here we give an overview of the method, compare the results to multi-dimensional simulations and investigate the effects of the progenitor and the equation of state on black hole formation.}

\section{Introduction}
\begin{figure}[tbh]
\centering

\begin{minipage}{.4\textwidth}
  \centering
  \includegraphics[width=1.05\linewidth]{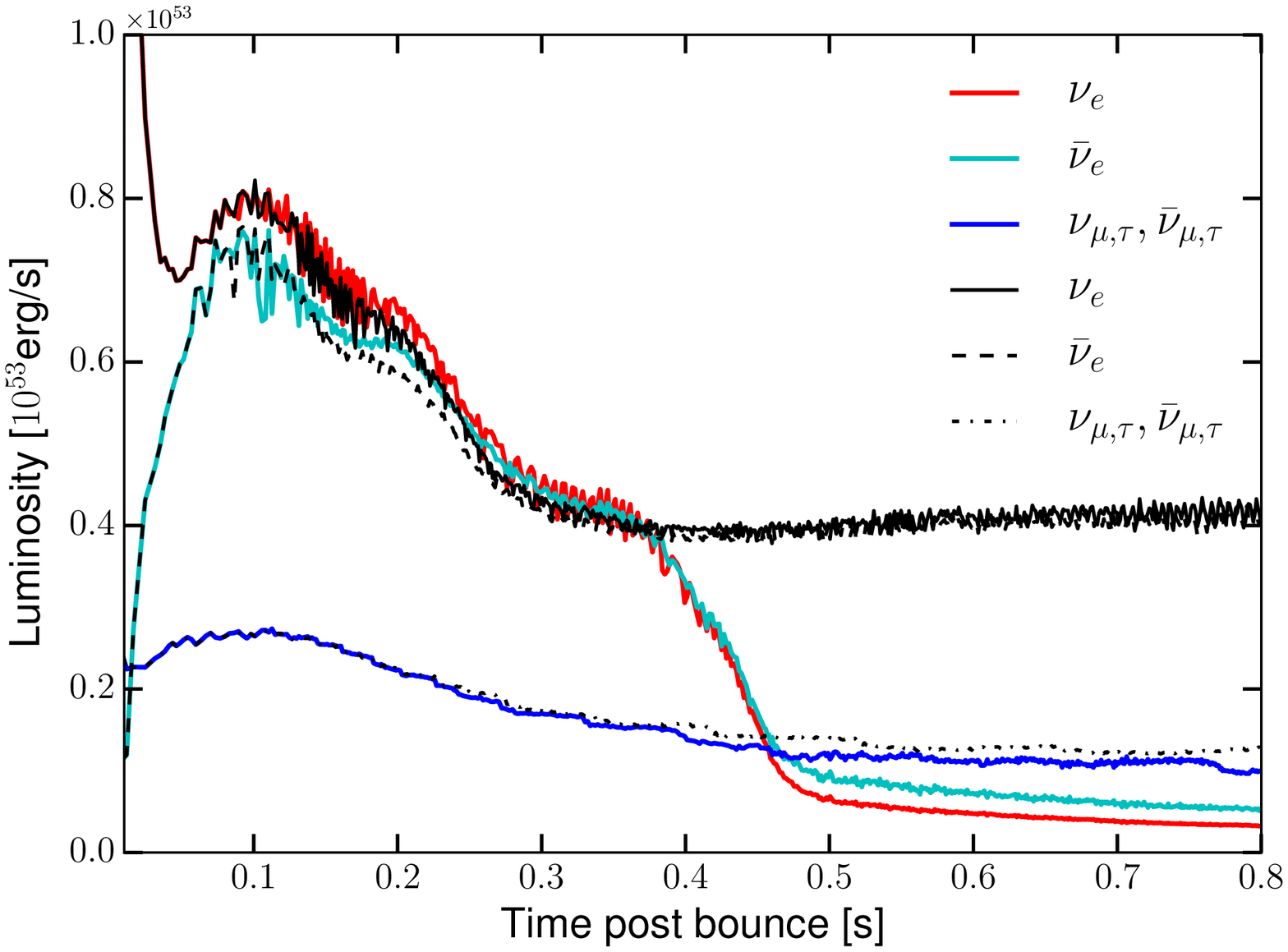}

  \caption{Neutrino luminosities for simulations with (colored lines) and without PUSH (black lines).}
    \label{lum}
\end{minipage}
\hspace{2cm}
\begin{minipage}{.4\textwidth}
  \centering
  \includegraphics[width=\linewidth]{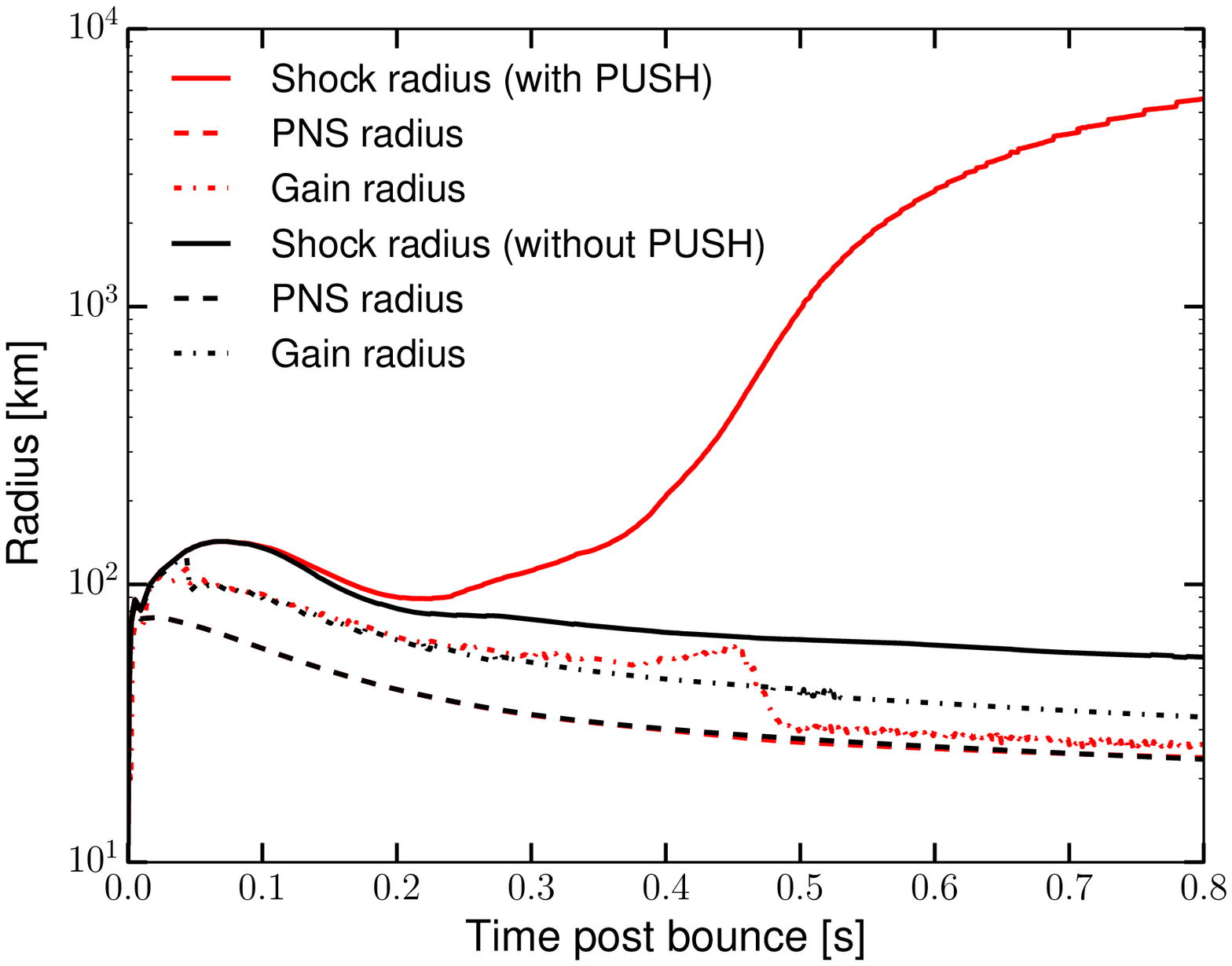}
 
  \caption{Temporal evolution of the shock radius, the PNS radius and the gain radius of a 20 M$_{\odot}$ progenitor (WH07 \cite{prog1}).}
  \label{radii}

\end{minipage}

\end{figure}

Core-collapse supernovae (CCSNe) occur at the end of the evolution of massive stars and the ejecta of these violent events contribute to the chemical evolution of the universe. The explosion mechanism of CCSNe is still not fully understood and self consistent one-dimensional simulations of CCSNe, including general relativity and detailed neutrino transport, do not lead to explosions, with the exception of the lowest-mass CCSN progenitors\cite{lowmass}.
Even though multi-dimensional simulations are promising to explode and well suited to investigate the explosion mechanism they are computationally too expensive to explore a large set of progenitors.
The presented parametrized one-dimensional framework (PUSH, introduced in\cite{push}) is well
suited to study explosive nucleosynthesis and remnant properties of a broad range of CCSN progenitors and with this also get a better understanding of phenomenon itself.
Spherically symmetric simulations show a smaller heating efficiency of electron neutrinos behind the shock due to an absence of convective motion. PUSH provides extra energy deposition in the heating region by tapping the energy of $\mu$- and $\tau$-  (anti)-neutrinos in otherwise consistent spherically symmetric simulations to mimic multi-dimensional effects (e.g., convection, SASI) that enhance neutrino heating. This enables a consistent evolution of the PNS and treatment of the electron fraction of the ejecta. Furthermore, after the onset of explosion the method also prevents a too strong decrease in  $\nu$-heating behind the shock due to a drop in electron (anti)neutrino luminosity that occurs in 1D simulations due to drastic reduction of the mass accretion rate on the PNS (see Figure \ref{lum}). Figure \ref{radii} shows the temporal evolution of the shock, gain and PNS radius with and without PUSH of a CCSN simulation of a 20 M$_{\odot}$ star.

\section{Comparison with multi-dimensional simulations}
\begin{figure}[tbh]
\centering
\begin{minipage}{.4\textwidth}
  \centering
  \includegraphics[width=\linewidth]{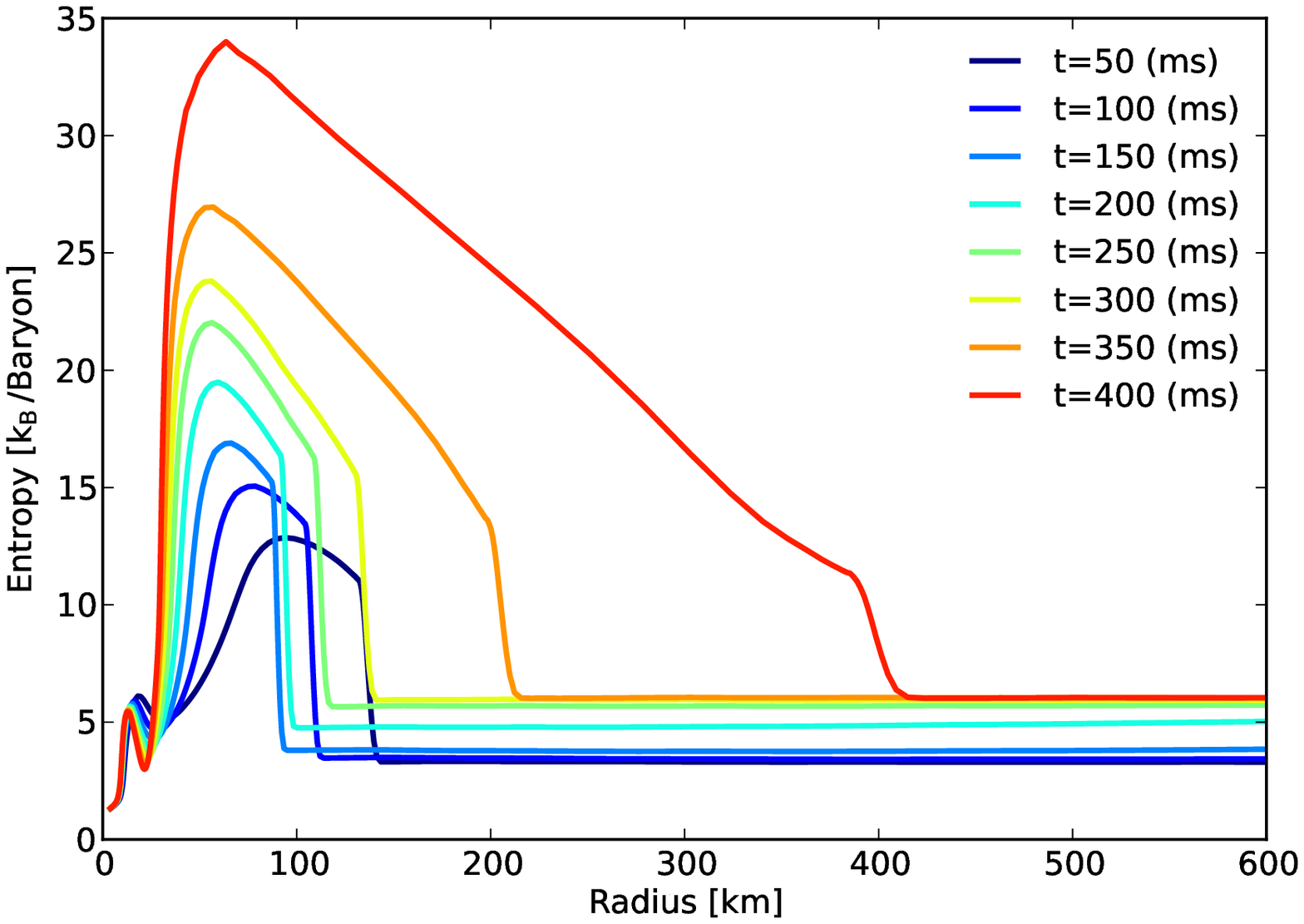}
  \caption{Spherically averaged entropy per baryon as a function of radius obtained from a 1D simulation with PUSH of a 20 M$_{\odot}$ progenitor.}
  \label{pushentropy}
\end{minipage}
\hspace{2cm}
\begin{minipage}{.4\textwidth}
  \centering
  \includegraphics[width=\linewidth]{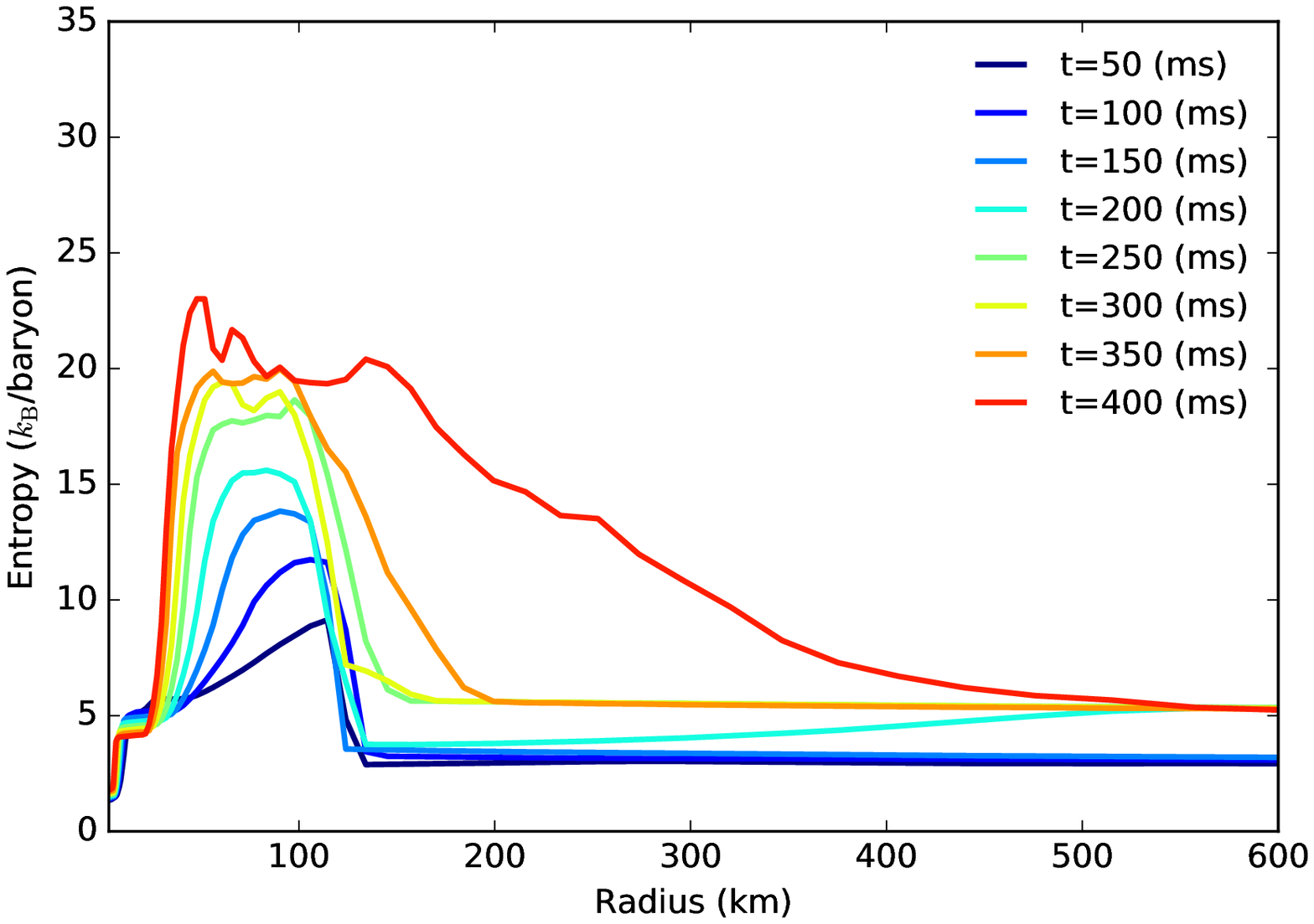}
  \caption{Spherically averaged entropy per baryon as a function of radius obtained from a 2D Flash simulation of a 20 M$_{\odot}$ progenitor.}
    \label{flashentropy}
\end{minipage}

\end{figure}

Overall PUSH shows a behaviour more consistent with multi-dimensional models than older methods (e.g. pistons and thermal bombs \cite{piston},\cite{thiel96}). Figures \ref{pushentropy} and \ref{flashentropy} show the spherically averaged entropy per baryon as a function of radius obtained from a 2D Flash simulation (see \cite{kcpan} and references therein) and from a 1D simulation with PUSH for the same progenitor and electron (anti)neutrino transport \cite{lieb1}. The comparison of the two figures shows on average a similar heating pattern. Such a comparison can be used as a further fit requirement (besides explosion energy and nucleosynthesis yields \cite{push}, see also proceeding of S. Sinha, this volume) for the free parameters of the PUSH method.

  \vspace{-1cm}

\section{Progenitor and Equation of State Dependence of Black Hole Formation}

To disentangle aspects - other than the explosion mechanism - that have an influence on black hole formation we investigate the effect that different choices of the equation of state and of the progenitor profiles can have in our 1D simulations. 

\begin{figure}
\centering
\begin{minipage}{.4\textwidth}
  \centering
  \includegraphics[width=\linewidth]{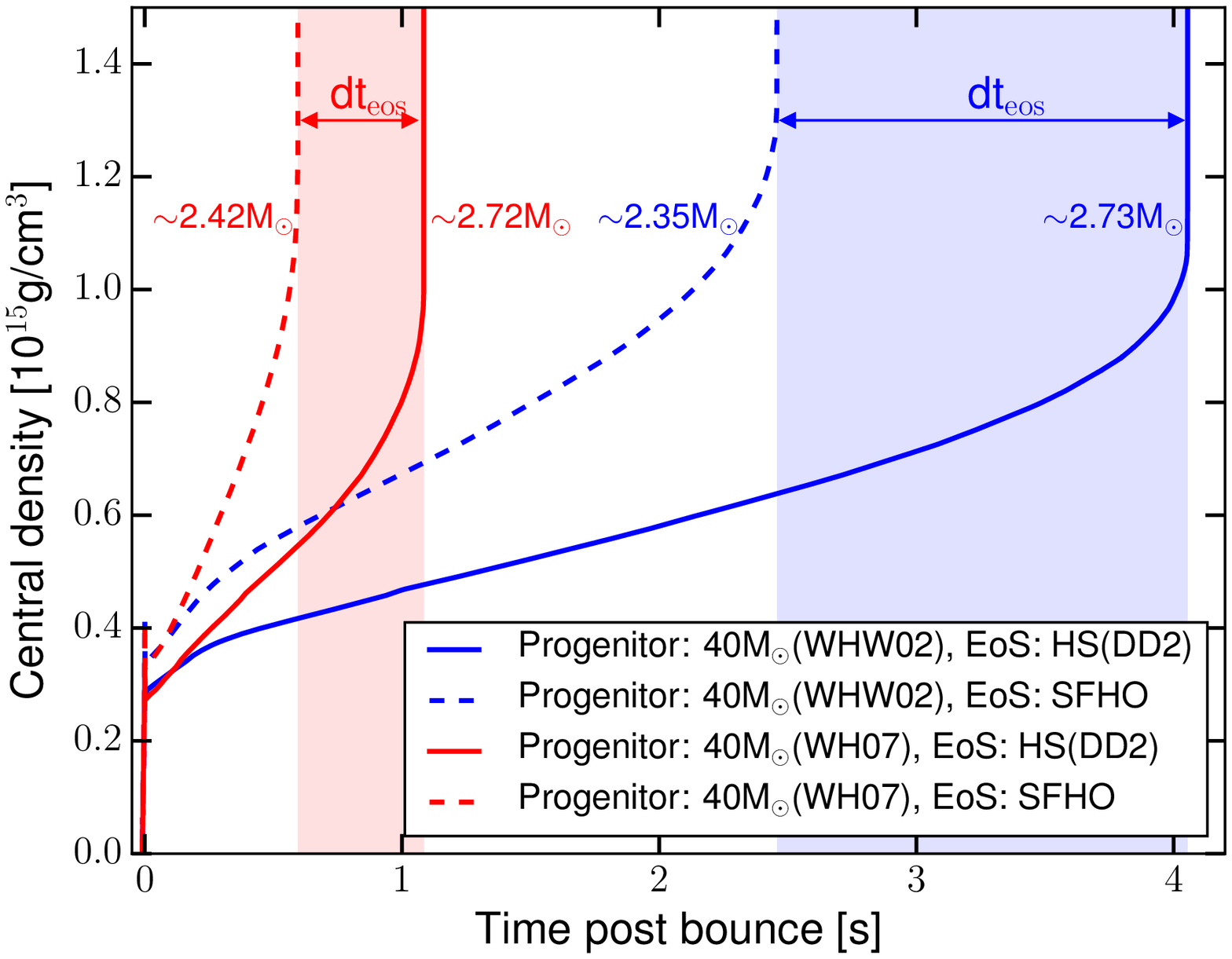}
  \caption{Temporal evolution of the central density of a 40 M$_{\odot}$ solar metallicity star for two progenitor models (WH07\cite{prog1} in red and WHW02\cite{prog3} in blue) and two equations of state (HS(DD2) solid lines, SFHO dashed lines, \cite{hempel},\cite{fischer}).}
  \label{fig:detailcollapse}
\end{minipage}
\hspace{2cm}
\begin{minipage}{.4\textwidth}
  \centering
  \includegraphics[width=\linewidth]{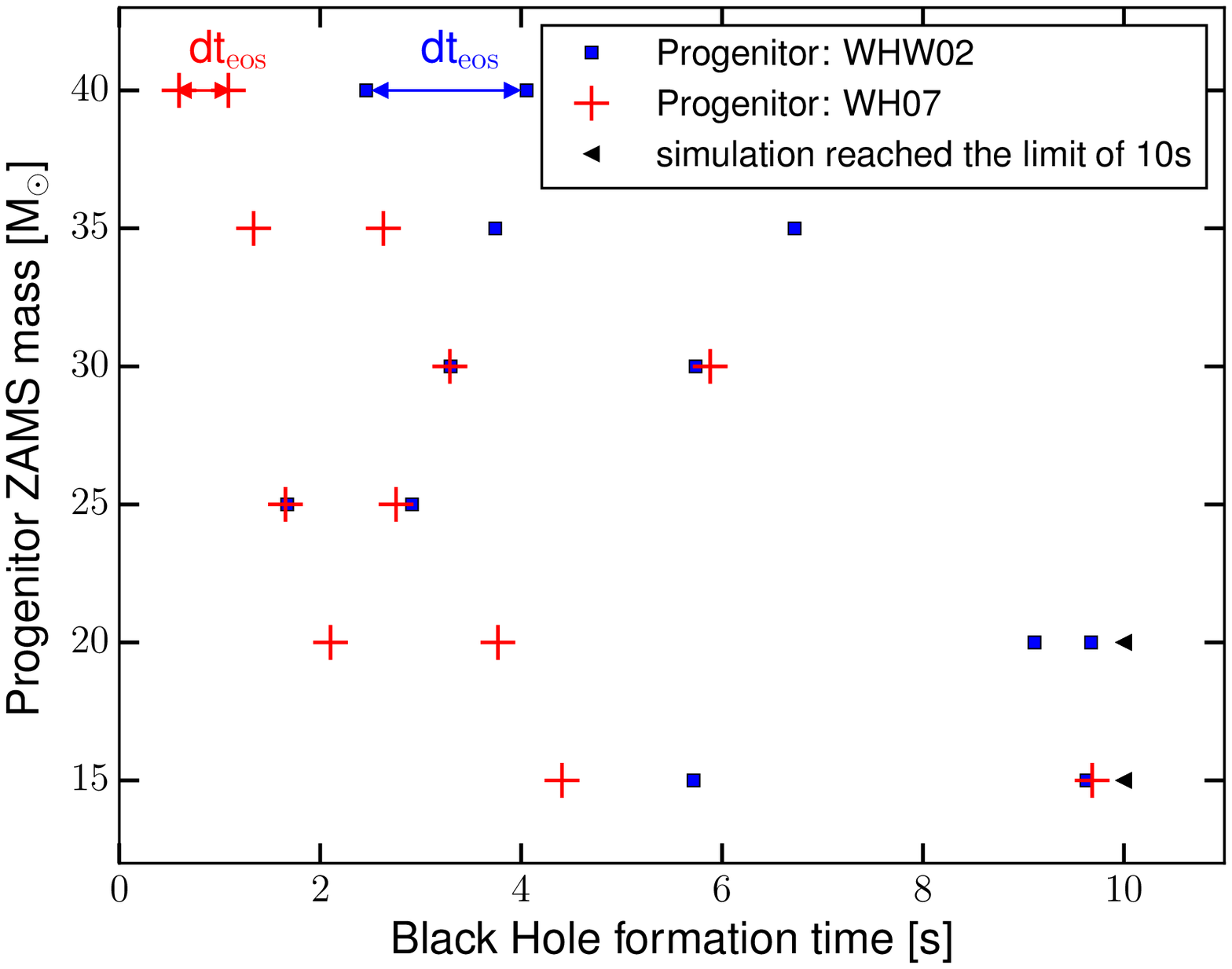}

  \caption{Black hole formation times for a collection of different progenitor ZAMS masses from two different progenitor sets (WH07\cite{prog1} in red and WHW02\cite{prog3} in blue).}
   \label{fig:overviewcollapse}
\end{minipage}
\end{figure}

Figure \ref{fig:detailcollapse} shows the temporal evolution of the central density of a 40 M$_{\odot}$ solar metallicity star for two progenitor models (WH07\cite{prog1} in red and WHW02\cite{prog3} in blue) and two equations of state (HS(DD2) solid lines, SFHO dashed lines, \cite{hempel},\cite{fischer},\cite{sfho}). The dependence of the black hole formation time on the equation of state (indicated by the colored areas) and the even stronger dependence on the progenitor model  for this progenitor ZAMS mass (difference between red and blue lines) is evident. Baryonic PNS masses at collapse are given next to the corresponding central density curves. In Figure \ref{fig:overviewcollapse} the black hole formation times for a set of different progenitor ZAMS masses are given. The differences for black hole formation time between the progenitors can be related to different accretion rates, which are correlated to compactness  $\xi_{M}=\frac{M/M_{\odot}}{R(M)/1000km}$.

\section{Conclusions and Outlook}

In comparison to traditional effective methods, as pistons or thermal bombs, PUSH is better suited to study explosive nucleosynthesis, especially of the innermost ejecta, due to the inclusion of more neutrino physics and the preservation of charged current reactions.  We have shown that the entorpy profiles obained with PUSH are similar to the spherical averages of multi-dimensional models and demonstrated the big effect the choice of progenitor and equation of state can have on black hole formation and thus on a study of explodability. It is planned to investigate the explodability of different progenitor sets with different equations of state with PUSH in the future.

\end{document}